\documentclass[12pt,a4paper]{article}
\usepackage{bbold}
\usepackage{graphicx,color}
\usepackage{bm,amsmath,amssymb}
\usepackage{pdfsync}
\usepackage{enumerate,float}
\usepackage{eurosym}

\pdfoutput=1

\newcommand\eq[1]{\begin{eqnarray}#1\end{eqnarray}}

\newcommand{\diff}{\mathrm{d}}

\date{}
\title{EASY HOLOGRAPHY}

\author{Heikki Arponen\footnote{\texttt{heikki.a.arponen@gmail.com}. Addr: Espoontie 12 A 2, 02770 Espoo, FINLAND}}
\begin{document}

\maketitle
\begin{center}
{\normalsize \emph{Essay written for the Gravity
Research Foundation 2013 Awards for Essays on Gravitation}}\\
{\small Submitted on 2013/03/31}
\end{center}

\abstract{It is argued that the role of infinite dimensional asymptotic symmetry groups in gravity theories are essential for a holographic description of gravity and therefore to a resolution of the black hole information paradox. I present a simple toy model in two dimensional hyperbolic/ anti-de Sitter (AdS) space and describe, by very elementary considerations, how the asymptotic symmetry group is responsible for the entropy area law. Similar results apply also in three dimensional AdS space. The failure of the approach in higher dimensional AdS spaces is explained and resolved by considering other asymptotically noncompact homogeneous spaces.}
%


%
\tableofcontents
%

\newpage

\section{Gravity in two dimensions}

Suppose that there is a theory of gravity in $\mathbb{R}^2$. It doesn't matter what the theory is, just that the symmetry group is $Diff(\mathbb{R}^2)$ (the group of diffeomorphisms that map $\mathbb{R}^2$ to itself) as dictated by the principle of general coordinate invariance. We consider a toy model with two cases of boundary conditions for gravity: first "free" boundary condition which doesn't restrict the symmetry group and then an asymptotically hyperbolic/anti-de Sitter boundary condition, which does. 

\subsection{Ink Model}

\begin{figure}[hb]
  \centering
  \def\svgwidth{0.8\textwidth}
  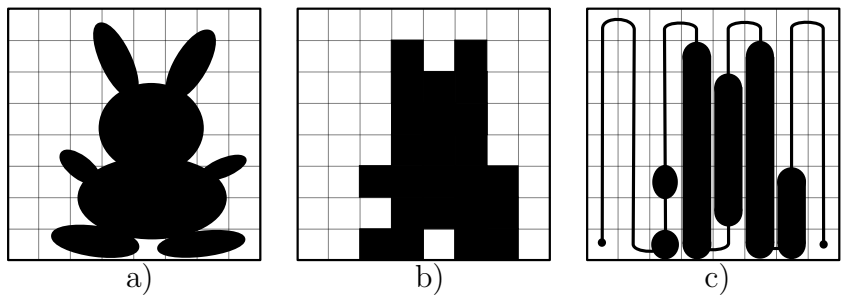
  \caption{a) A random ink blot on an $8\times 8$ lattice. b) The same blot after coarse graining. c) Any coarse grained ink blot can be obtained by deforming any other ink blot by e.g. a conformal mapping on $\mathbb R^2$.\label{coarsediffeo}}
\end{figure}

The "ink model" is defined in three steps. First, splash some ink on paper so that it forms a simply connected region (for simplicity). Then place a grid, or a lattice, on top of it. Any kind of a lattice will do, but we can work with a standard evenly spaced square lattice. As a third step, define a "coarse graining" map, similarly to what is used in the renormalization group, which colors a lattice square black if most of the square area is covered in the ink, and leaves it white otherwise. We call the resulting image a \emph{configuration} of the ink model. The process is illustrated in Fig. \ref{coarsediffeo}.\\

A symmetry group is such that an action by a group element transforms a solution into another one. Suppose now that the symmetry group $Diff(\mathbb{R}^2)$ is large enough so that it can be used to obtain any (topologically equivalent) solution by acting on a known solution, e.g. the flat metric. Then all solutions, or configurations, can be labelled with elements in $Diff(\mathbb{R}^2)$. Now we can ask the following question: \emph{how many distinct configurations are there in the ink model when we consider all the transformations in $Diff(\mathbb{R}^2)$ on the initial state?}\\

It is easy to find the answer by heuristic arguments. The diffeomorphism group can be used to morph the ink blot into pretty much any desired shape, so one could just as well start by mapping the blot into a ball that sits in the middle of the lower left square. It is then a simple process to obtain any desired configuration (even ones that have holes in the coarse grained image) by stretching the blot through all the squares and thinning at the squares that are supposed to be white after the coarse graining, see Fig. \ref{coarsediffeo} c) for an example.\footnote{In fact, it would be possible to do this with only a conformal transformation by resorting to the Riemann mapping theorem.} Therefore the number of configurations, or degrees of freedom, for an $N$ by $N$ lattice is $2^{N^2}$. The relative weights of the configurations are simply assumed to be equal, so that the entropy of the ink model is just the logarithm of the above number, $S = N^2 \log 2$. If the length of a square is $\epsilon$ and the length of the side of the system is denoted as $L=\epsilon N$, then $S = (L/\epsilon)^2 \log 2$. The entropy is therefore proportional to the volume of the system (which of course in this two dimensional case is the area), as one might expect.\\

\subsection{Hyperbolic Ink Model}

Consider now the same (or any) two dimensional gravity theory but with a different boundary condition. Specifically, the system is now enclosed in the upper half plane $\mathbb{H}^2 \doteq \left\{ (x, y ) \in \mathbb{R}^2 |y>0  \right\}$ with the requirement that the metric is \emph{asymptotically hyperbolic} (or anti-de Sitter) near the boundary, i.e. that the metric tends asymptotically to the metric of the hyperbolic space,
\eq{\diff s^2 = \frac{1}{y^2} \left(\diff x^2 + \diff y^2  \right).\label{metric}}
This reduces the group of diffeomorphisms into a subgroup of \emph{asymptotic symmetries} which leave the above metric asymptotically invariant as the boundary at $y=0$ is approached. It is in fact convenient to express the metric in complex coordinates with $z = x+ i y$ as
\eq{\diff s^2 = \frac{\diff z \diff \bar z}{\Im(z)^2}.}
It is then easy to see that the asymptotic symmetries can be expressed as conformal maps from the upper half plane to itself: $\left(z , \bar z\right) \to \left(f(z) , f(\bar z)\right)$.\footnote{We also demand that the $f$ are isotopic to identity.} The corresponding conformal group will here be denoted as $Conf\! \left(\mathbb{H}^2 \right)$. Note that at the boundary $y=0$ the group action becomes simply $x \to f(x)$. These transformations form the diffeomorphism group of the circle\footnote{The upper half plane is conformally equivalent to a unit disc, i.e. the boundary is actually $S^1$.}, which means that there is a local isomorphism $Conf\! \left(\mathbb{H}^2 \right) \simeq Diff(S^1)$. Now we can ask a similar question as in the $Diff$ invariant case: \emph{how many distinct configurations are there in the ink model when we consider all the transformations in $Conf\! \left(\mathbb{H}^2 \right) \subset Diff(\mathbb{R}^2)$ on the initial state?}\\

I claim that for large $N$ the number of states is proportional to $2^N$, which leads to an area law for the entropy, $S \propto L$. The claim can be proved easily by a scaling argument and some elementary algebra.\\

\begin{figure}[h]
  \centering
  \def\svgwidth{0.3\textwidth}
  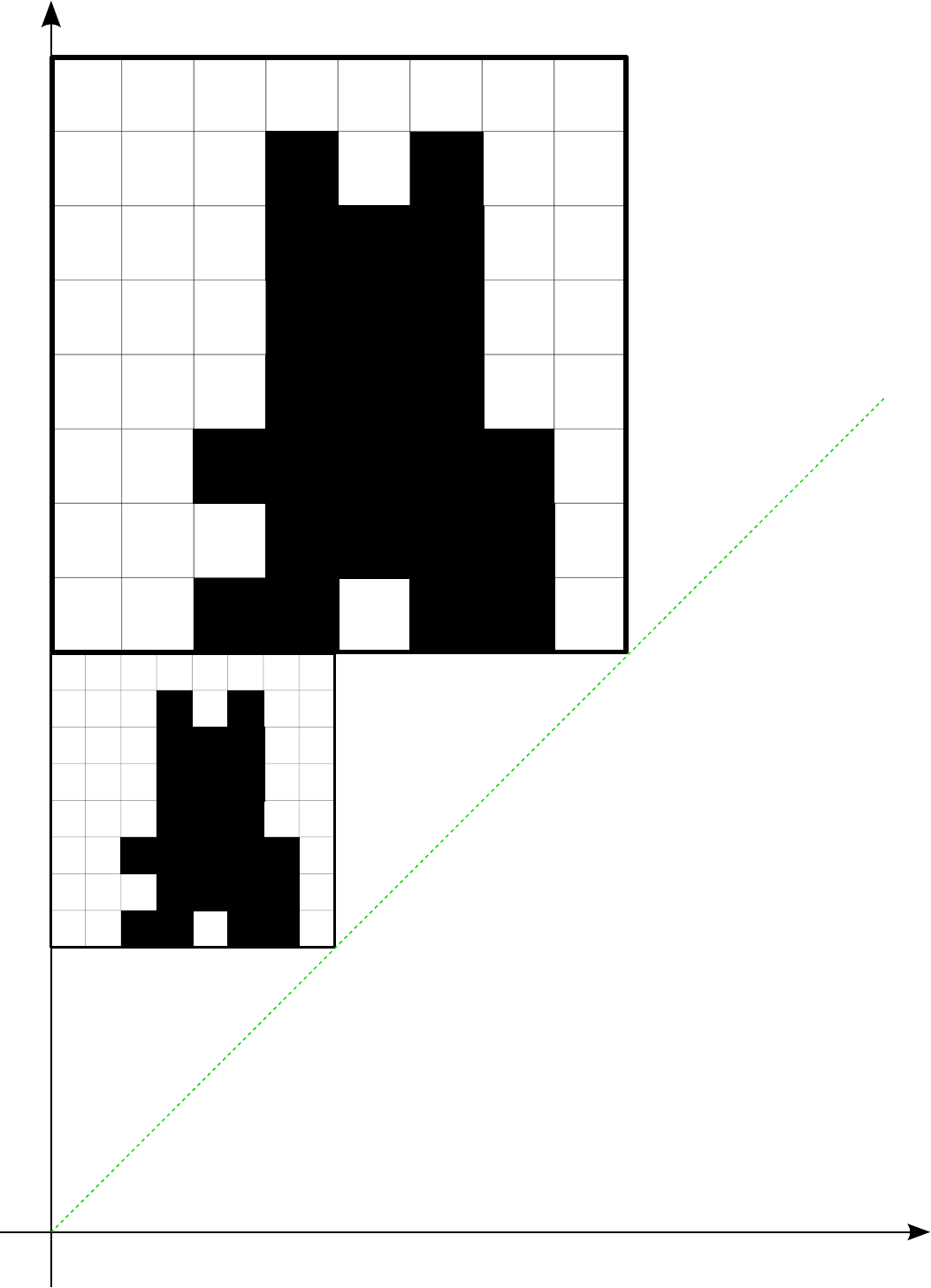
  \caption{An $8\times 8$ lattice scaled by a factor of $2$. The systems are equivalent, since scaling is an isometry.\label{scaledbunny}}
\end{figure}

\begin{figure}[h]
  \centering
  \def\svgwidth{0.9\textwidth}
  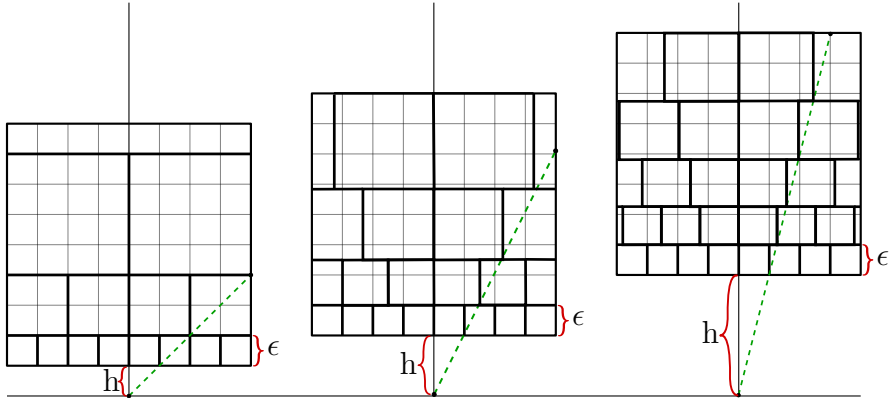
  \caption{True degrees of freedom for an $8 \times 8$ lattice when increasing the height $h$ from the boundary. Each of the hyperbolic squares inside a lattice have the same number of degrees of freedom.\label{truedof}}
\end{figure}

Consider e.g. an $8\times 8$ lattice as in Fig. \ref{scaledbunny}. A scaling $z \to \lambda z$ for $\lambda \in \mathbb R$ is an isometry and therefore a symmetry of the system. Then the system obtained by scaling the original system by, say, a factor of $2$ is identical to the original one and, specifically, has the same number of configurations. This is true also \emph{inside} a system, which leads us to conclude that the true degrees of freedom reside on a hyperbolic tiling of the system. This is illustrated in Fig. \ref{truedof} with varying distance $h$ from the boundary. One should specifically note that the number of hyperbolic squares approaches the number of original squares as $h$ approaches infinity.\\

Counting the number of configurations proceeds as follows: Consider an $N \times N$ lattice of square side length $\epsilon$ placed at height $h$ from the boundary, as in Fig. \ref{truedof}. Then the next row of hyperbolic lattices starts at height $h+\epsilon = \left( \frac{h+\epsilon}{h}\right) h$ and the next one at $\left( \frac{h+\epsilon}{h}\right)^2 h$ and so on, until the $N$:th row is reached. In general the squares won't all fit inside the lattice, in which case we simply ignore the "residual" parts: for large $N$ the error will be small. The number of rows of the hyperbolic lattice is then (approximately)
\eq{N' = \frac{\log \left( 1+ N \delta \right)}{\log \left( 1+ \delta \right)},} 
where $\delta \doteq \epsilon / h$. The approximate number of total hyperbolic squares is then
\eq{\nu(\delta,N)\doteq  N \sum\limits_{k=0}^{N'-1} \left( 1+ \delta \right)^{-k} = \frac{1+\delta}{1+N \delta} N^2.}

There are three obvious limiting cases. The first is large $N$ with
\eq{\nu(\delta,N) \approx (1+h/\epsilon) N}
resulting in the entropy 

\eq{S \doteq \log 2^{\nu(\delta,N)} \approx  (1+h/\epsilon) N \log 2.} 

The second case is when $\delta  \to \infty$ (i.e. $h \to 0$ if $\epsilon$ is thought of as the Planck constant), which gives simply $S = N \log 2$. The third case is $\delta \to 0$, i.e. $h \to \infty$, which leads to $S = N^2 \log 2$, i.e. the entropy respects the usual volume law of classical thermodynamics. We conclude that the entropy respects an area law near the boundary, but tends to the usual thermodynamic entropy far from the boundary. It is interesting to think about this in the context of the Riemann Mapping Theorem, which is here seen to fail near the boundary, i.e. there are simply connected regions that cannot be mapped to each other. The reason for this is that we have only "one half" of the conformal mappings at hand, since the upper half plane maps to itself.\\

The derivation of the entropy area law above relied on two important facts: 

\begin{itemize}
\item[1)] Existence of an infinite dimensional asymptotic symmetry group. Indeed, if the asymptotic symmetries were only finite dimensional, the entropy thus derived would behave as $\propto \log N$. 
\item[2)] Non-existence of a \emph{too} infinite dimensional asymptotic symmetry group, such as $Diff(\mathbb R^2)$. This would result in a volume law for the entropy and therefore to the black hole information paradox.
\end{itemize}

\section{$AdS_3$ and higher dimensions}

The asymptotically $AdS_3$ case is an almost trivial generalization of the $AdS_2$ case. There is now an extra time-like coordinate and the metric is
\eq{\diff s^2 = \frac{1}{y^2}\left( \diff x^2 - \diff t^2 + \diff y^2 \right).}
The asymptotic symmetry group is simply the pseudo-conformal group of two dimensional Minkowski space and it is locally isomorphic to $Diff(S^1) \times Diff(S^1)$. For a constant time slice, the entropy area law is identical to the above case of $AdS_2$/hyperbolic space.\\

Then how about higher dimensions? For $\mathbb H^2$, a.k.a. $AdS_2$, the asymptotic symmetry group is (locally) isomorphic to $Diff(S^1)$, for $AdS_3$ it is $Diff(S^1) \times Diff(S^1)$ and for $AdS_4$ it turns out to be... $SO(2,2;\mathbb{R})$. Damn! As already mentioned, in this case the "ink model" construction fails and the entropy derived by the above methods can only be at most logarithmic in $N$.\\

But not so fast! When looking at the $AdS_{d}$ spaces at 2, 3 and 4 dimensions carefully, we note something odd. In a sense, $AdS_4$ is \emph{not} the logical step from 2 and 3 dimensions. In fact, one can define the $AdS_{d}$ spaces as the homogeneous spaces $SO(2,d-1)/SO(2,d-2)$.\footnote{...over the real numbers.} In low dimensions there happens to exist so-called "accidental isomorphisms" of Lie groups (or rather their covering groups), namely $SO(2,1) \simeq SL(2)$ and $SO(2,2) \simeq SL(2) \times SL(2)$. We note then that $AdS_3$ space is just the Lie group $SL(2)$ and $AdS_2$ is the homogeneous (symmetric) space $SL(2)/SO(2)$! It was the current author's idea to then find out what kind of asymptotic symmetries the five dimensional symmetric space $SL(3)/SO(3)$ might possess.\cite{arponen} It turns out that the asymptotic symmetries indeed do form an infinite dimensional symmetry group, which, according to a soon to be published work by the current author et al., is $Diff(S^2)$.

\section{Concluding remarks}

It is tempting to draw the conclusion that the resolution of the black hole information paradox requires a holographic universe with a specific infinite dimensional asymptotic symmetry group, which can only result from a noncompact geometry of spacetime. In other words, the principle of general coordinate invariance would need to be replaced with a "principle of asymptotic coordinate invariance", which would mean that not all diffeomorphisms can be symmetries in gravity, but only the ones that preserve the asymptotic geometry. Of course it would make sense to try to find a \emph{four} dimensional space-time with an infinite dimensional asymptotic symmetry group, which sounds like an interesting research project.

\bibliography{bibsit3}

\begin{thebibliography}{1}

\bibitem{arponen}
Heikki Arponen.
\newblock Infinite symmetry on the boundary of $sl(3)/so(3)$.
\newblock {\em J. Math. Phys.}, 53(3):033512, 2012.

\end{thebibliography}
\bibliographystyle{plain}

\end{document}